\documentclass[twocolumn,superscriptaddress,amssymb,nobibnotes,aps,prb]{revtex4-1}

\usepackage{graphicx}
\usepackage{color}
\usepackage{amsfonts}
\usepackage{amsmath}
\usepackage{graphicx}
\usepackage{bm}
\usepackage{amssymb}
\usepackage{dcolumn}
\usepackage{cases}
\usepackage{booktabs}
\usepackage{setspace}
\usepackage{array}
\usepackage{float}
\usepackage{placeins}
\usepackage{blindtext}
\usepackage{physics}
\usepackage{upgreek}

\usepackage{array}
\usepackage{float}
\usepackage{blindtext}
\newcommand{\singlefigwidthlarge}{0.48}

\makeatletter
    \renewcommand\@make@capt@title[2]{%
     \@ifx@empty\float@link{\@firstofone}{\expandafter\href\expandafter{\float@link}}%
      {\textbf{#1}}\@caption@fignum@sep#2\quad}%
    \makeatother
   
\makeatletter 
\renewcommand{\fnum@figure}{\textbf{Figure~\thefigure}}
\makeatother

\begin{document}

\title{Thickness dependence of the anomalous Hall effect in thin films of the topological semimetal Co$_2$MnGa}

\author{Anastasios Markou}
\email{Anastasios.Markou@cpfs.mpg.de}
\affiliation{Max Planck Institute for Chemical Physics of Solids, N\"othnitzer Str. 40, 01187 Dresden, Germany}
\author{Dominik Kriegner}
\affiliation{Max Planck Institute for Chemical Physics of Solids, N\"othnitzer Str. 40, 01187 Dresden, Germany}
\author{Jacob Gayles}
\affiliation{Max Planck Institute for Chemical Physics of Solids, N\"othnitzer Str. 40, 01187 Dresden, Germany}
\author{Liguo Zhang}
\affiliation{Max Planck Institute for Chemical Physics of Solids, N\"othnitzer Str. 40, 01187 Dresden, Germany}
\author{Yi-Cheng Chen}
\affiliation{Max Planck Institute for Chemical Physics of Solids, N\"othnitzer Str. 40, 01187 Dresden, Germany}
\affiliation{Department of Materials Science and Engineering, National Chiao Tung University, 30010 Hsinchu, Taiwan}
\author{Benedikt Ernst}
\affiliation{Max Planck Institute for Chemical Physics of Solids, N\"othnitzer Str. 40, 01187 Dresden, Germany}
\author{Yu-Hong Lai}
\affiliation{Department of Materials Science and Engineering, National Chiao Tung University, 30010 Hsinchu, Taiwan}
\author{Walter Schnelle}
\affiliation{Max Planck Institute for Chemical Physics of Solids, N\"othnitzer Str. 40, 01187 Dresden, Germany}
\author{Ying-Hao Chu}
\affiliation{Department of Materials Science and Engineering, National Chiao Tung University, 30010 Hsinchu, Taiwan}
\author{Yan Sun}
\affiliation{Max Planck Institute for Chemical Physics of Solids, N\"othnitzer Str. 40, 01187 Dresden, Germany}
\author{Claudia Felser}
\affiliation{Max Planck Institute for Chemical Physics of Solids, N\"othnitzer Str. 40, 01187 Dresden, Germany}

%%%%%%%%%%%%%%%%%%%%%%%%%%%%%%%%%%%%%%%%%%%%%%%%%%%%%%%%%%%%%%%%%%%%%%

\date{\today}

\begin{abstract}
	
 Topological magnetic semimetals promise large Berry curvature through the distribution of the topological Weyl nodes or nodal lines and further novel physics with exotic transport phenomena. We present a systematic study of the structural and magnetotransport properties of Co$_2$MnGa films from thin (20\,nm) to bulk like behavior (80\,nm), in order to understand the underlying mechanisms and the role on the topology. The magnetron sputtered Co$_2$MnGa films are \textit{L}$2_{\mathrm {1}}$-ordered showing very good heteroepitaxy and a strain-induced tetragonal distortion. The anomalous Hall conductivity was found to be maximum at a value of 1138\,S/cm, with a corresponding anomalous Hall angle of 13\,\%, which is comparatively larger than topologically trivial metals. There is a good agreement between the theoretical calculations and the Hall conductivity observed for the 80\,nm film, which suggest that the effect is intrinsic. Thus, the Co$_2$MnGa compound manifests as a promising material towards topologically-driven spintronic applications. 

\end{abstract}

\maketitle

%%%%%%%%%%%%%%%%%%%%%%%%%%%%%%%%%%%%%%%%%%%%%%%%%%%%%%%%%%%%%%%%%%%%%%
\section{Introduction}

The experimental realization of graphene, characterized by a low energy Dirac dispersion,~\cite{CastroNeto2009b} reinvigorated the interest in utilizing and understanding the topological electronic nature of materials. One such avenue led to the topological insulators determined by the surface Dirac dispersion states and gapped states in the bulk.~\cite{Hasan2010} Contrary to topological insulators, the exotic topological semimetals are advantageous due to an intimate correlation of the dispersion of the bulk and surface states. The Dirac,~\cite{Young2012,Wang2012} Weyl,~\cite{Burkov2011,Weng2015} and nodal line~\cite{Burkov2011a} semimetals are the most well-known examples. The Weyl semimetal (WSM) is found in systems with a lack of inversion symmetry or time reversal symmetry, where two momentum-space distributed Weyl nodes of opposite chirality form and are linked by the so-called Fermi-arc surface bands.~\cite{Wan2011,Sun2015,Batabyal2016,Inoue2016} The finite Berry curvature associated with the Weyl nodes or gapped nodal line leads to exotic phenomena, such as the chiral anomaly,~\cite{Son2013,Huang2015,Zhang2016,Parameswaran2014} magnetoptical~\cite{Higo_2018} and transport responses,~\cite{Manna_2018a} the anomalous Hall effect (AHE),~\cite{Nagaosa2010,Burkov2014} and the Nernst effect counterparts.~\cite{Noky2018a,Noky2018,Guin2019} In magnetic Heusler compounds, this has become a major point of interest,~\cite{Manna_2018a} where large values of the AHE have been associated with the momentum-space distribution of the Weyl nodes and gapped nodal lines. 

Recently, Wang \textit{et al.}~\cite{Wang2016} proposed that several Co-based full Heusler compounds realize WSMs. They focused on Co$_2$ZrSn compound and found that two Weyl points exist close to the Fermi energy when the magnetization is along the [110] direction. Chang \textit{et al.}~\cite{Chang2016} pinpointed the topological semimetal states in the Co$_2$TiZ (Z = Si, Ge or Sn) compounds by first-principle calculations. K{\"u}bler and Felser~\cite{Kubler_2016} suggested that the experimentally observed large AHE in Co$_2$MnAl films~\cite{Vilanova2011} is due to the distribution of four Weyl points just above the Fermi edge, while they, also, predicted similar results for the Co$_2$MnGa compound. Further, it was found to be a nodal line and Weyl semimetal.~\cite{Belobourdas2017}

The bulk Co$_2$MnGa Heusler compound crystallizes in the cubic Cu$_2$MnAl-type structure with space group~\textit{Fm$\bar{3}$m}\ (No.\,225) and lattice constant  \textit{a}\,=\,5.77\,\AA. The Co atoms occupy the Wyckoff position 8\textit{c}, whereas the Mn and Ga occupy the positions 4\textit{b} and 4\textit{a}, respectively. It is a semimetallic ferromagnet with a large saturated magnetic moment of $\textit{M}_{\mathrm s}$ \,=\,4.05\,\textit{$\mu$}$_{\mathrm B}$/f.u., following the Slater Pauling rule,~\cite{Galanakis2002} and high Curie temperature of $\textit{T}_{\mathrm C}$\,=\,694\,K.~\cite{Webster1971, Brown2000} Single crystals of Co$_2$MnGa were found to exhibit large AHE~\cite{Manna_2018} and anomalous Nernst effect (ANE),~\cite{Sakai_2018,Guin2019} which originates from the large Berry curvature distribution around Fermi energy and associated with nodal lines~\cite{Belobourdas2017,Chang_2017} or Weyl points.~\cite{Kubler_2016} Albeit, for device applications a proper understanding of the thin film limit is required. There have been some studies on thin films of Co$_2$MnGa, in particular, a MgO/Co$_2$MnGa/Pd stack with perpendicular magnetic anisotropy that shows AHE~\cite{Ludbrook_2017} and Co$_2$MnGa films with one of the largest values reported for the ANE.~\cite{Reichlova_2018} To fully utilize the topological properties, one must understand the dependence of the transport properties from the bulk-like films to the thin film limit. In the thin film limit, the properties of the surface play a significant role due to the intrinsic topology rooted in the electronic structure.

In this work, we present the structural, magnetic and transport properties of high-quality Co$_2$MnGa films with thickness ranging from 20 to 80\,nm. We performed systematic x-ray diffraction (XRD), transmission electron microscopy (TEM), magnetic and transport characterization of films heteroepitaxially grown on MgO substrates. We find that the magnetotransport properties vary with the temperature and the film thickness. A strong AHE is observed in the 80\,nm thick film that agrees with bulk first-principle calculations of the intrinsic AHE. Lastly, the 80\,nm film shows to be the upper limit for both the AHE (1138\,S/cm) and anomalous Hall angle (AHA) of 13\%, while the 20\,nm film displays an AHE of 840\,S/cm and an AHA of 10\%.

%%%%%%%%%%%%%%%%%%%%%%%%%%%%%%%%%%%%%%%%%%%%%%%%%%%%%%%%%%%%%%%%%%%%%%

\section{Experimental details}
 
 Co$_2$MnGa films with thicknesses of 20, 40, 60 and 80\,nm have been grown heteroepitaxially on MgO\,(001) single crystal substrates. A BESTEC~UHV magnetron sputtering system was used for the deposition of the films, with Co\,(5.08\,cm), Mn\,(5.08\,cm) and Mn$_{50}$Ga$_{50}$\,(5.08\,cm) sources in confocal geometry. The target to substrate distance was 20\,cm. Prior to deposition, the chamber was evacuated to a base pressure less than 8\,$\times$\,10$^{-9}$\,mbar, while the process gas (Ar 5\,N) pressure was 3\,$\times$\,10$^{-3}$\,mbar. The Co$_2$MnGa films were grown by co-sputtering and the individual sputter rates were adjusted to obtain the desired composition. The Co was deposited by applying 34\,W dc power, the Mn by applying 6\,W dc power and the Mn$_{50}$Ga$_{50}$ by applying 22\,W dc, and the total rate was 0.58\,\AA/s.  The substrates were rotated during deposition, to ensure homogeneous growth. The films were grown at 550\,$^{\circ}$C and then post-annealed $in\,situ$ for an additional 20 minutes to improve the chemical ordering. All samples were capped with a 3\,nm thick Al film at room temperature to prevent oxidation.
 
     \begin{figure}
 	\centering
 	\includegraphics[width=\singlefigwidthlarge\textwidth]{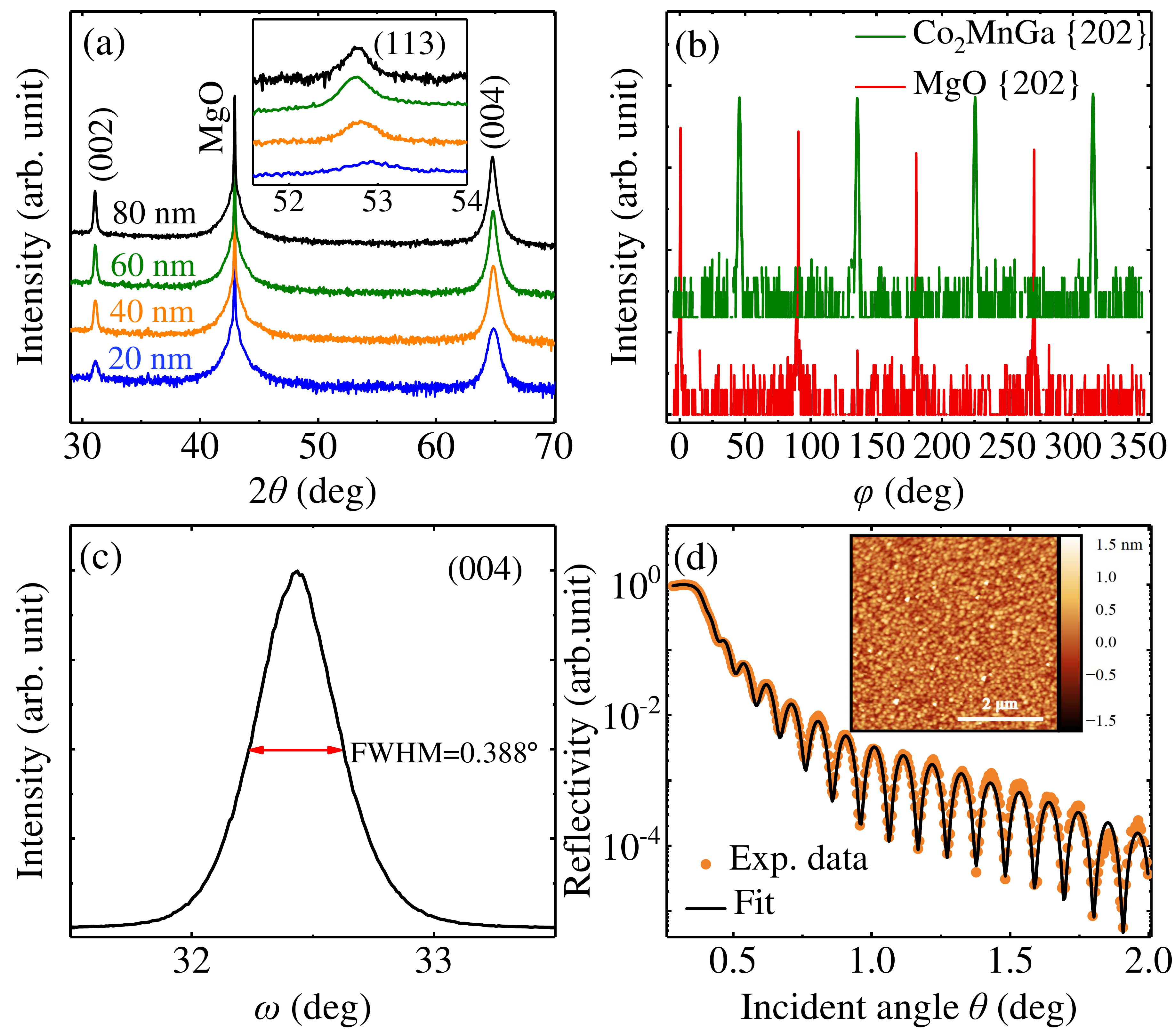}
 	\caption{(a) XRD pattern of the Co$_2$MnGa films with different thickness (20-80\,nm). The inset shows the (113) assymetric reflections. (b) $\varphi$-scan patterns of the \{202\} planes from the 60\,nm Co$_2$MnGa film and  the MgO substrate. (c) Rocking curve ($\omega$-scan) of (004) reflection from a 80\,nm Co$_2$MnGa. (d) X-ray reflectivity pattern of the 40\,nm Co$_2$MnGa, where the solid line represents the least-squares fit to the data.}
 	\label{fig:XRD}
 \end{figure}

 Stoichiometry was estimated as Co$_{51}$Mn$_{25}$Ga$_{24}$ by energy-dispersive x-ray spectroscopy (EDXS) with an experimental uncertainty of 2\,\%. X-ray diffraction (XRD) and x-ray reflectivity (XRR) were measured with a Panalytical X$'$Pert$^3$ MRD diffractometer, using Cu–K$_{\alpha1}$ radiation ($\lambda$\,=\,1.5406\,\AA). The growth rates and the film thicknesses were determined by a quartz crystal microbalance and confirmed by using XRR measurements. The atomic force microscopy (AFM) image was obtained with a MFP-3D Origin$^+$ microscope from Oxford Instruments Asylum Research. High-resolution scanning transmission electron microscopy (HRSTEM) was performed using a JEOL ARM200F microscope operated at 200\,kV.  Additionally, the JEOL was equipped with EDXS for element mapping with high spatial resolution. Diffraction patterns were collected by a FEI Tecnai operated at 200\,kV.  Cross-section samples were prepared by focused ion beam milling (FIB). A protective C-Pt layer was deposited on the stack before starting the cross-section preparation. Magnetization measurements were carried out using a Quantum Design (MPMS3 SQUID-VSM) magnetometer and the transport measurements were carried out on bars with a rectangular shape (7\,mm\,x\,2\,mm) in a six-probe method using bonded Al wires contacts with low-frequency alternating current (ETO, PPMS9 Quantum Design). 
 
%%%%%%%%%%%%%%%%%%%%%%%%%%%%%%%%%%%%%%%%%%%%%%%%%%%%%%%%%%%%%%%%%%%%%%

\section{RESULTS AND DISCUSSION}

\subsection{Structural properties}

Different x-ray scattering measurements, $2\theta$-$\omega$ scan, rocking curve ($\omega$-scan) and sample azimuth ($\varphi$) scan, were performed to study the structure, the crystallinity and the heteroepitaxial relationship between the films and the substrate, respectively. The in-plane lattice mismatch between the 45$^{\circ}$ rotated MgO unit cell ($\sqrt{2}$\,$a_{\mathrm {MgO}}$) and bulk Co$_2$MnGa is 3.2\,\%, which allows the heteroepitaxial growth of (001) oriented films. Fig.~\ref{fig:XRD}(a) shows the XRD patterns of the 20, 40, 60 and 80\,nm Co$_2$MnGa films. In addition to the (002) reflection of the MgO substrate, all the samples exhibit exclusively the (002) and (004) reflections of the cubic Co$_2$MnGa, indicating (001)-oriented films. 

\begin{table}
	\caption{Lattice parameters, volume, tetragonallity $c/a$ and the full width at half maximum (FWHM) of the (004) rocking curve profile for Co$_2$MnGa thin films with different thickness.}
	\begin{ruledtabular}
		\begin{tabular}{cccccc}
			Thickness & $c$ & $a$ & Volume & $c/a$ & FWHM \\ 
			(nm) & (\AA) & (\AA) & (\AA$^3$) & & (deg)\\
			\hline
			20 & 5.728 & 5.811 & 193.4 & 0.986 & 0.495 \\
			40 & 5.741 & 5.793 & 192.7 & 0.991 & 0.485 \\
			60 & 5.751 & 5.785 & 192.4 & 0.994 & 0.424 \\
			80 & 5.747 & 5.786 & 192.4 & 0.993 & 0.388 \\
		\end{tabular}
	\end{ruledtabular}
	\label{tab:structure}
\end{table}

The properties of the Heusler compounds are strongly dependent on the occupation of the crystallographic sites. To determine if the films are fully chemically ordered in \textit{L}$2_{\mathrm {1}}$-type, in addition to (002) reflection, the presence of the superstructure (111) or (113) reflections are needed.  The inset in Fig.~\ref{fig:XRD}(a) shows $2\theta$-$\omega$ scans of the asymmetric (113) Bragg reflection of the Co$_2$MnGa thin films. The data were acquired in coplanar diffraction geometry with a linear detector and the acquisition software integrates along the rocking angle ($\omega$). Quantitative analysis of the integrated intensities for all Bragg reflections, considering the diffraction geometry, as well as polarization, Lorentz, and absorption corrections \cite{Kriegner2013} consistently showed that below 10\% of \textit{B}2-type disorder is present in the films, which therefore can be considered almost fully chemically ordered.

$\varphi$-scan patterns of the \{202\} planes from the Co$_2$MnGa film and the MgO substrate are depicted in Fig.~\ref{fig:XRD}(b). The reflections of Co$_2$MnGa show four-fold symmetry with 90$^{\circ}$ intervals, suggesting single crystalline epilayers with well-defined in-plane orientation. By comparing the diffractions of film and substrate a 45$^{\circ}$ in-plane rotation of the Co$_2$MnGa unit cell is observed with respect to the MgO substrate. The crystallographic orientation relationship is thus determined as Co$_2$MnGa(001)[110]$\|$MgO(001)[100].

The crystal quality of the Co$_2$MnGa films was evaluated from the FWHM values of rocking curves measured around (004) reflection. The  small FWHM\,$\approx$\,0.388$^{\circ}$ (Fig.~\ref{fig:XRD}(c)) suggests that the 80\,nm film shows high crystalline quality with low mosaicity. The FWHM remains below 0.5$^{\circ}$ for the thinner films indicating a very good crystal quality (Table~\ref{tab:structure}).

 \begin{figure}
	\centering
	\includegraphics[width=\singlefigwidthlarge\textwidth]{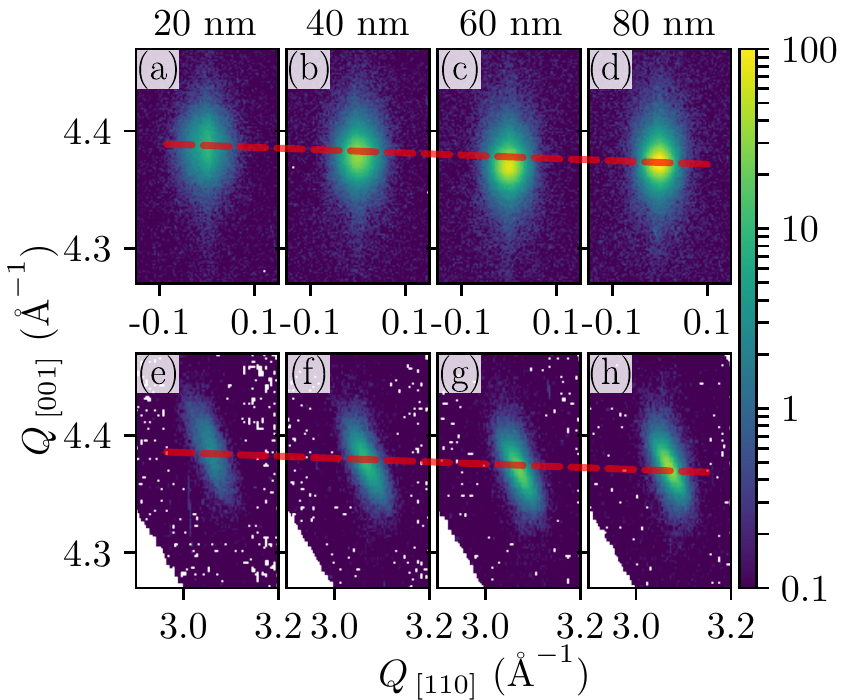}
	\caption{X-ray diffraction reciprocal space maps of Co$_2$MnGa thin films. Panels (a)-(e) and (f)-(j) show the diffracted intensity on a logarithmic color scale in the vicinity of the (004) and (224) Bragg peaks of the films, respectively. The $x$ and $y$ axis of the plot show the momentum transfer $Q$ along the [110] and and [001] directions defined as $Q_{[110]} = 4\pi/\lambda \sin\theta \sin(\omega-\theta)$, and $Q_{[001]} = 4\pi/\lambda \sin\theta \cos(\omega-\theta)$. $\lambda$\,=\,1.5406\,\AA~is the wavelength of the x-ray radiation. Red lines are guides to the eye and show that with increasing film thickness the peak position is changing.}
	\label{fig:RSM}
\end{figure}

In spite of the high-temperature growth, all the films are very smooth. Fig.~\ref{fig:XRD}(d) shows the XRR pattern of the 40\,nm Co$_2$MnGa film with 3\,nm Al capping layer. The thickness oscillations that start after the critical angle up to 2 deg and beyond indicate smooth films and sharp interfaces. From the good agreement between the experimental data (data points) and the model curve calculated using a modified Parrat formalism~\cite{Pietsch2004} (solid line), we deduce the structural parameters, such as thickness, roughness, and density. Thickness and density agree within 2\% with the nominal values and the determined substrate/film and film/capping layer roughnesses are found to be 3.4 and 4.5\,\AA, respectively. The smooth surface topography is confirmed with AFM from a 40\,nm thick film without Al capping layer, as depicted in the inset of Fig.~\ref{fig:XRD}(d). The film is continuous and very smooth with a root mean square roughness of 3\,\AA, which is in good agreement with the XRR measurement.

In order to understand how the strain-induced by the substrate influences the Co$_2$MnGa films, we performed reciprocal space map measurements around the (004) and (224) Bragg reflections of Co$_2$MnGa. Fig.~\ref{fig:RSM} shows that for all film thicknesses, the defined peaks are observed for both Bragg peaks, which indicates the epitaxial growth for all thicknesses. The peak position varies, which reflects a change of the lattice parameters with the film thickness. The extracted lattice parameters are summarized in Table~\ref{tab:structure} and show that the in-plane lattice parameter slightly decreases with thickness, while simultaneously the out-of-plane parameter increases, hence the unit cell volume remains almost constant. The thinner films display a tetragonal distortion, while thicker films approach the crystalline cubic structure found in bulk compounds. Given the reported lattice parameters of our films in Table I, we conclude that the effective symmetry of the films is reduced to space group ~\textit{I}$4$/\textit{mmm} (No.\,139).

%%%%%%%%%%%%%%%%%%%%%%%%%%%%%%%%%%%%%%%%%%%%%%%%%%%%%%%%%%%%%%%%%%%%%%%%%%%%%
\subsection{TEM investigation}

\begin{figure}
	\centering
	\includegraphics[width=\singlefigwidthlarge\textwidth]{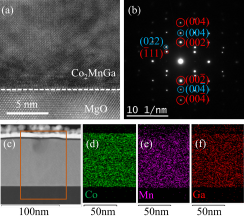}
	\caption{(a) Cross-section HRSTEM image of the 80\,nm Co$_2$MnGa film grown on MgO~ substrate. (b) SAED pattern showing the diffraction spots from  Co$_2$MnGa (blue open circles) and MgO (red open circles). (c) Cross-section HAADF-STEM image, where the orange box denotes the area, where chemical mapping was performed. Elemental mapping of (d) Co (green), (e) Mn (purple) and (f) Ga (red).}
	\label{fig:TEM}
\end{figure}

TEM was performed in the 80\,nm thick Co$_2$MnGa to evaluate the film quality on the nanoscale. In Fig.~\ref{fig:TEM}(a) we show the cross-section HRSTEM image of the film. The high-quality growth of the Co$_2$MnGa film on MgO substrate is manifested by the crystal lattice of the Co$_2$MnGa film. Despite the lattice mismatch of 2.9\,\%  between film and substrate, the Co$_2$MnGa epilayer is characterized by good heteroepitaxy with no significant defects. Furthermore, the interface between film and substrate is very sharp with atomic-level flatness.

We depict the selected area electron diffraction (SAED) pattern of the same sample in Fig.~\ref{fig:TEM}(b), where the electron beam is parallel to the [110] zone axis of the MgO substrate. The red and blue open circles correspond to the diffraction spots from the MgO substrate and the Co$_2$MnGa film, respectively. The two different sets of diffraction spots are aligned, confirming the heteroepitaxial growth of the epilayer on the substrate.

Fig.~\ref{fig:TEM}(c) shows the high-angle annular dark-field (HAADF) STEM image, where from bottom to top the MgO substrate, the Co$_2$MnGa film and the Al capping layer are clearly shown in different brightness. The film is continuous, smooth and with the expected thickness of 80\,nm. The element distribution within the Co$_2$MnGa films was analyzed by EDXS element mapping and the orange box in the HAADF-STEM image (Fig.~\ref{fig:TEM}(c)) indicates the area where the elemental analysis was carried out. The spatial distribution of the count rate intensity of the Co, Mn, and Ga elements are represented with different colors in Fig.~\ref{fig:TEM}(d)-(f). All elements were detected at exactly the same sample regions, therefore confirming the homogeneity of the Co$_2$MnGa.

\subsection{Magnetic and transport properties}
\begin{figure}
	\centering
	\includegraphics[width=\singlefigwidthlarge\textwidth]{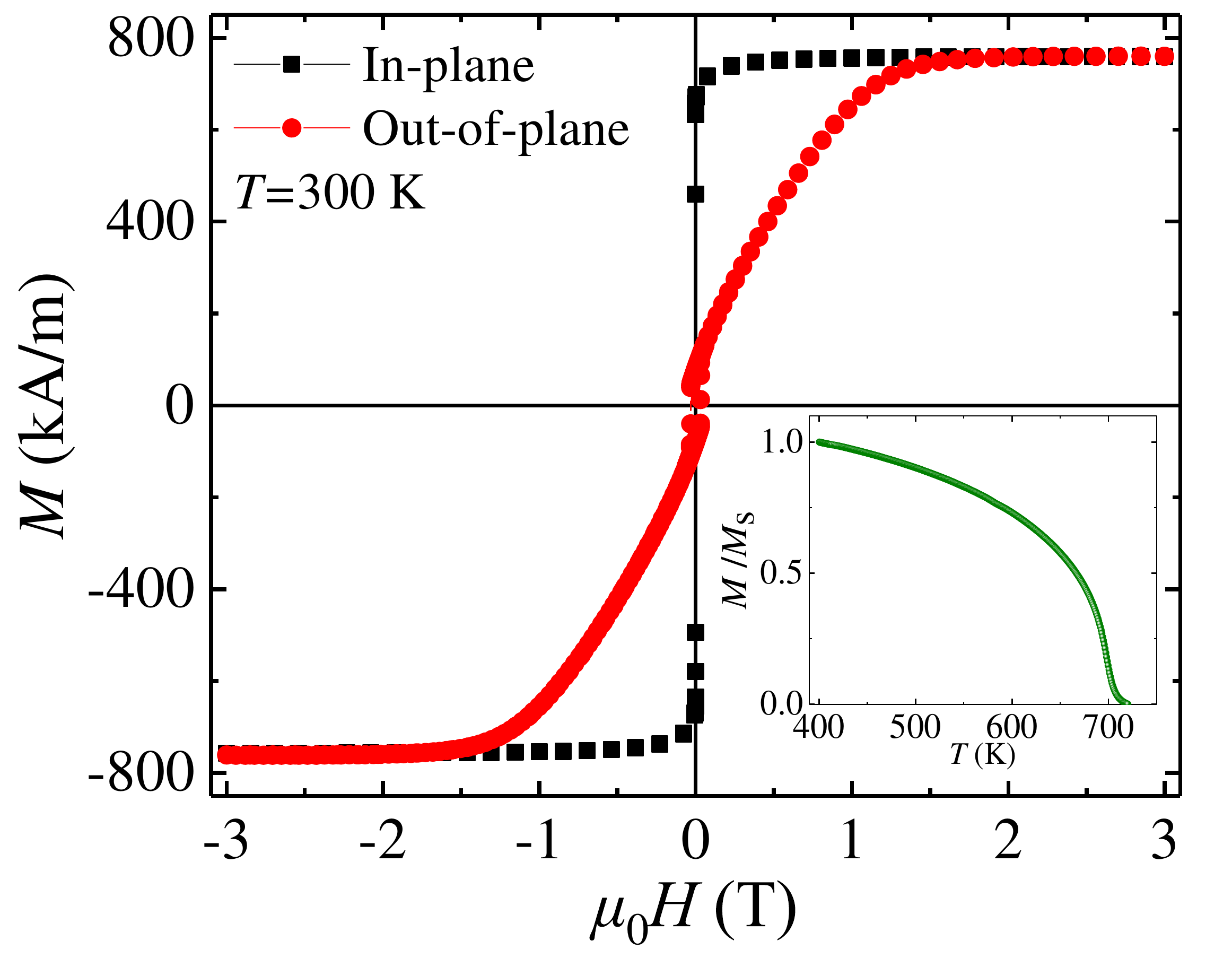}
	\caption{In-plane and out-of-plane magnetization hysteresis loops of the 80\,nm Co$_2$MnGa film. The inset shows the temperature dependence of normalized magnetization for the 60\,nm Co$_2$MnGa film.}
	\label{fig:hysteresis}
\end{figure}

In Fig.~\ref{fig:hysteresis} we show typical in-plane and out-of-plane magnetization hysteresis loops for the 80\,nm Co$_2$MnGa film measured at 300\,K. The Co$_2$MnGa film is a soft magnet with an in-plane magnetic easy axis. The saturation magnetization is $\textit{M}_{\mathrm s}$\,=\,760\,kA/m and the coercivity is \textit{$\mu$$_0$}$\textit{H}_{\mathrm c}$\,=\,0.5\,mT. The Curie temperature was measured using SQUID magnetometry with an oven option. The magnetization was recorded with a constant in-plane field of \textit{$\mu$$_0$}$\textit{H}$\,=\,50\,mT while warming up the sample, as shown in the inset of Fig.~\ref{fig:hysteresis}. The curve is normalized at 400\,K and the extracted Curie temperature is 700 $\pm$ 5\,K. The saturation magnetization and the Curie temperature ($\textit{M}_{\mathrm s}$ \,=\,4.20\,\textit{$\mu$}$_{\mathrm B}$/f.u. and $\textit{T}_{\mathrm C}$\,=\,700\,K)are slightly higher compared to the bulk~\cite{Webster1971} ($\textit{M}_{\mathrm s}$ \,=\,4.05\,\textit{$\mu$}$_{\mathrm B}$/f.u. and $\textit{T}_{\mathrm C}$\,=\,694\,K) and the calculated~\cite{Galanakis2002} ($\textit{M}_{\mathrm s}$ \,=\,4.058\,\textit{$\mu$}$_{\mathrm B}$/f.u.) values. This can be attributed to the small amount of disorder and/or off-stoichiometry in our films and it is also in a good agreement with electronic structure calculations, which take into account disorder in Mn/Ga sites.~\cite{Galanakis2006}   

\begin{figure*}
	\centering
	\includegraphics[width=\textwidth]{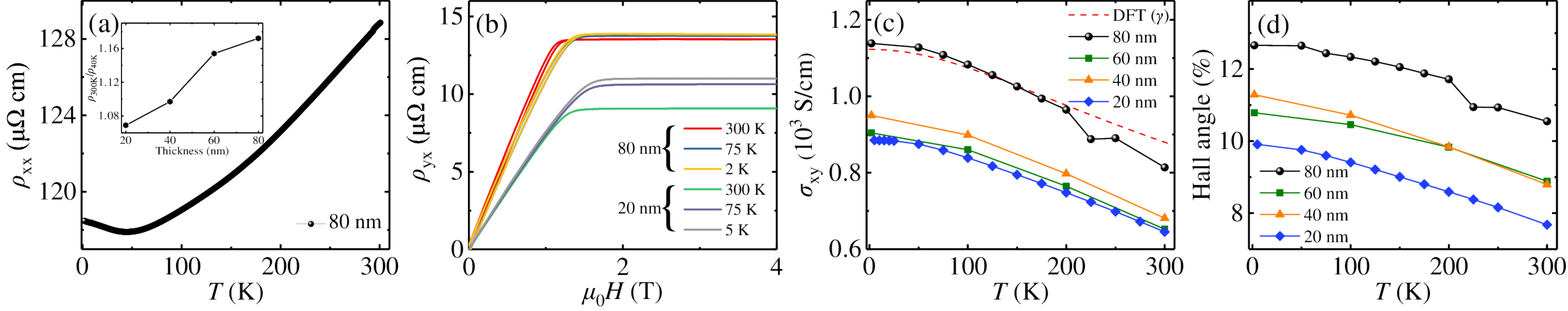}
	\caption{(a) Temperature dependent longitudinal resistivity of the 80\,nm Co$_2$MnGa film. The inset shows the thickness dependent RRR. (b) Hall resistivity as a function of the applied magnetic field for the 80\,nm and 20\,nm Co$_2$MnGa films collected at 300, 75, 5 and 2\,K. (c) Temperature dependent Hall conductivity of Co$_2$MnGa films with different thickness. The red dashed line represents the constant $\gamma$ broadening from {\it ab-initio} calculations. (d) Temperature dependent anomalous Hall angle of Co$_2$MnGa films with different thickness.}
	\label{fig:transport}
\end{figure*}

We show the transport properties of the Co$_2$MnGa films in Fig.~\ref{fig:transport}. The transport measurements are collected with the magnetic field and the current applied along the [001] and [110] directions of the Co$_2$MnGa, respectively. Fig.~\ref{fig:transport}(a) depicts the temperature dependent resistivity at zero field for the 80\,nm Co$_2$MnGa film. The film shows metallic behavior with a residual resistivity of 110\,$\upmu$\textit{$\Upomega$}\,cm at 40\,K. Above 45\,K, $\textit{$\rho$}_{\mathrm {xx}}$(\textit{T}) is nonlinear, which indicates mixed phonon and magnon scattering state.~\cite{Goodings1963} The upturn of longitudinal resistivity below 45\,K is attributed to defects or due to diffusion channel electron-electron interaction mechanism as recently reported in \textit{L}$2_{\mathrm {1}}$-Co$_2$MnAl films.~\cite{Zhu_2017} The inset in Fig.~\ref{fig:transport}(a) shows the residual resistivity ratio (RRR\,=\,$\textit{$\rho$}_{\mathrm {300\,K}}$\,/\,$\textit{$\rho$}_{\mathrm {40\,K}}$) as a function of film thickness. The RRR increases linearly as the thickness increases and reaches 1.17 for the 80\,nm film, which compares well with other Co-based full Heusler compound films.~\cite{Vidal_2011,Gabor_2015,Zhu_2017}

In Fig.~\ref{fig:transport}(b) the first quadrant of the Hall resistivity loops are shown for the 80\,nm and the 20\,nm at $T$\,=\,300, 75, 5 and 2 K. We analyze the Hall resistivity in the saturated state by using
\begin{equation}\label{eq:1}
\rho_{\mathrm {yx}}\,=\,\textit{R}_{\mathrm 0}\textit{$\mu$$_0$}\textit{H}\,+\,\rho_{\mathrm {yx}}^{\mathrm{AHE}},
\end{equation}
where \textit{$\mu$$_0$}, $\textit{R}_{\mathrm 0}$ and $\rho_{\mathrm {yx}}^{\mathrm{AHE}}\,=\,\textit{R}_{\mathrm S}\textit{M}$ are the permeability of free space, the ordinary Hall coefficient, $\textit{R}_{\mathrm S}$ is the anomalous Hall coefficient and \textit{M} is the magnetization perpendicular to the plane of the film, respectively. There is a clear trend that shows a saturated Hall resistivity above applied fields of 2\,T for all thicknesses and temperatures. The tetragonal distortion of the Co$_2$MnGa films is also reflected in the Hall resistivity, where the 80\,nm film, which approaches the cubic structure, saturates close to an applied field of 1.2\,T and the more tetragonally distorted 20\,nm film at 1.5\,T. Furthermore, the magnitude of the AHE varies with the thickness and the temperature, where it is maximized at 2\,K. The maximum and minimum of the Hall resistivity is found to be $\rho_{\mathrm {yx}}$\,=\,14\,$\upmu$\textit{$\Upomega$}\,cm, and 11\,$\upmu$\textit{$\Upomega$}\,cm for the 80\, and 20\,nm films, respectively. In order to compare our experimental results with theoretical calculations, we extract the Hall conductivity as a function of temperature.

The experimental values of the anomalous Hall conductivity ($\sigma_{\mathrm {xy}}^{\mathrm {AHE}}$) are estimated by the Hall resistivity $\textit{$\rho$}_{\mathrm {yx}}$ and the longitudinal resistivity as:
\begin{equation}\label{eq:2}
\sigma_{\mathrm {xy}}^{\mathrm {AHE}}\,=\,\frac{\rho_{\mathrm {yx}}}{\rho^2_{\mathrm {yx}}\,+\,\rho^2_{\mathrm {xx}}}. 
\end{equation}
In Fig~\ref{fig:transport}(c) the temperature dependence of $\sigma_{\mathrm {xy}}^{\mathrm {AHE}}$ for all films are compared to the theoretical calculations. For all films, the largest $\sigma_{\mathrm {xy}}^{\mathrm {AHE}}$ is observed at the lowest temperature, and is gradually reduced with increasing temperature. The thicker sample (80\,nm) shows a large value of $\sigma_{\mathrm {xy}}^{\mathrm {AHE}}$\,=\,1138\,S/cm at 2\,K and $\sigma_{\mathrm {xy}}^{\mathrm {AHE}}$\,=\,814\,S/cm at 300\,K, which are consistent with the bulk values,~\cite{Guin2019} while in thinner samples the $\sigma_{\mathrm {xy}}^{\mathrm {AHE}}$ is reduced. 

We calculate the AHE in the bulk Co$_2$MnGa compound using interpolated tight-binding Hamiltonian on a 351$^3$ {\bf k}-grid from the first-principle electronic structure converged on a 13$^3$ {\bf k}-grid from the {\small VASP} code.~\cite{Kresse1996} The AHE is calculated using the Kubo formula within the constant $\gamma$ broadening approximation:~\cite{Ryogo1957,Nagaosa2010}
\begin{flalign}\label{eq:3}
 \sigma_{ij}^{\mathrm {AHE}}  &\overset{\gamma \rightarrow 0 }{=}  \nonumber \\
&\frac{2e^2\hbar}{\mathcal{N}} \sum_{{\bf k},n}^\mathrm{occ}\sum_{m\neq n}\mathrm{Im}\left[\frac{\mel{\psi_{{\bf k},n}}{v_i}{\psi_{{\bf k},m}}\mel{\psi_{{\bf k},m}}{v_j}{\psi_{{\bf k},n}}}{(\mathcal{E}_{{\bf k},m}-\mathcal{E}_{{\bf k},n})^2+\gamma^2}\right].
\end{flalign}
Here, ${v_i}\,=\,\frac{1}{\hbar}\frac{\partial H}{\partial k_i}$ is the canonical velocity operator, and $\ket{\psi_{{\bf k},n}}$ is the eigenstate of $H$ with eigenvalue $\mathcal{E}_{{\bf k},n}$. The $\gamma$ dependence of the AHE in Fig.~\ref{fig:transport}(c) shows an excellent agreement with the experimental results of the 80\,nm film. This suggests that the 80\,nm film is sufficient to reproduce the bulk properties of Co$_2$MnGa and that the anomalous Hall conductivity is due to the intrinsic mechanisms. It is important to note that Eq.~\ref{eq:3} does not take into account side-jump and skew-scattering contributions to the AHE which can play a crucial role in films.~\cite{Lowitzer2010,Kovalev2010,Weischenberg2011} However, the experimental results show a decrease in the AHE with the decrease of the film thickness and with the same temperature dependence. This constant shift of the AHE with thickness variation suggest an increase of side-jump mechanism that has a similar temperature dependence as the intrinsic Berry curvature.~\cite{Nagaosa2010} The side-jump is opposite in sign to the intrinsic mechanism and may be due to the \textit{B}2-type disorder or the increased role of the surfaces as the thickness is decreased. Since, the \textit{B}2-type disorder is also present in bulk Co$_2$MnGa compounds,~\cite{Webster1971,Brown2000} therefore we attribute the side-jump to the increased role of the surfaces. The skew-scattering contribution nominally displays a strong variation with temperature, due to the linear dependence on the longitudinal resistivity.~\cite{Nagaosa2010} Furthermore, there may be phonon-assisted skew-scattering effects that show intrinsic behavior which cannot be disentangled in the current {\it ab-initio} theory.~\cite{Gorini2015a} The Weyl nodes and nodal lines are primarily due to the spin-up channel which leads to a suppression of spin scattering~\cite{Guin2019} analogous to the case in ferromagnetic half-metals. A recent work of MgO/Co$_2$MnGa/Pd stacks~\cite{Ludbrook_2017} shows longitudinal resistivity $\sim$\,30\,$\upmu$\textit{$\Upomega$}\,cm smaller and the AHE is an order of magnitude smaller (MgO/Co$_2$MnGa/Pd stack: $\sigma_{\mathrm {xy}}^{\mathrm {AHE}}$\,$\approx$\,125\,S/cm), compared to our films. This work claimed all three AHE mechanisms to be present, however dominated by the intrinsic and side-jump mechanisms.  

The anomalous Hall angle $\theta_{\mathrm {H}}$\,=\,$\sigma_{\mathrm {xy}}^{\mathrm {AHE}}\,/\,\sigma_{\mathrm {xx}}$ reflects the ability of a material to deviate the electron flow from the direction of the longitudinal electric field, due to the anomalous Hall effect. The temperature dependence of Hall angle for all Co$_2$MnGa films is shown in Fig.~\ref{fig:transport}(d). We find that the AHA varies with the temperature and the film thickness in a similar way as the temperature dependence of the Hall conductivity. The AHA of all films has a maximum at 2\,-5\,K and decreases with increasing temperature. A large AHA up to $\theta_{\mathrm {H}}$\,=\,12.7\,\% is observed at 2\,K for the 80\,nm film and remains high even at room temperature ($\theta_{\mathrm {H}}$\,=\,10.5\,\%). The combined large values of $\sigma_{\mathrm {xy}}^{\mathrm {AHE}}$ and $\theta_{\mathrm {H}}$  in our films suggest an intrinsic mechanism with large Berry curvature in a metallic topological semimetal. The $\sigma_{\mathrm {xy}}^{\mathrm {AHE}}$ and the $\theta_{\mathrm {H}}$ of the Co$_2$MnGa films are similar to Co$_3$Sn$_2$S$_2$,~\cite{Liu2018} with the gapped nodal line band structure~\cite{Xu2018}. The anomalous Hall conductivity displays a similar magnitude to metallic films such as \textit{L}$1_{\mathrm {0}}$-FePt,~\cite{Allen_1996} the noncollinear antiferromagnetic Weyl semimetals~\cite{Nakatsuji_2015,Nayak_2016} and the Dirac metal Fe$_3$Sn$_2$,~\cite{Ye_2018} however, the $\theta_{\mathrm {H}}$ is an order of magnitude larger, due to the unique topological electronic structure.

%%%%%%%%%%%%%%%%%%%%%%%%%%%%%%%%%%%%%%%%%%%%%%%%%%%%%%%%%%%%%%%%%%%%%%
\section{CONCLUSIONS}

In summary, we have studied the structural and magnetotransport properties of high-quality magnetron sputtered Co$_2$MnGa films with varying thickness. The Co$_2$MnGa films are (001)-oriented and almost fully-ordered in \textit{L}$2_{\mathrm {1}}$-type structure. The thinner films show a slight tetragonal distortion, whereas the thicker films approach the perfect cubic structure. Magnetic measurements reveal high magnetization and Curie temperature. We find that the magnetotransport properties vary with the temperature and the film thickness. The 80\,nm film shows a large anomalous Hall conductivity up to 1138\,S/cm, accompanied by a large anomalous Hall angle that is maximum at 13\,\%. The strong AHE signal in the 80\,nm thick film agrees very well with bulk first-principle calculations with a constant $\gamma$ band broadening approximation, which suggests that the experimentally observed AHE is due to intrinsic mechanism in the bulk limit, and as the thickness decreases the side-jump mechanism has a significant effect. Our work provides a pathway to develop thin film devices that include nodal line and Weyl topological properties intrinsic to the electronic structure.

\begin{acknowledgments}

\end{acknowledgments}
This work has been funded by  the ERC Advanced Grant No. 291472 \textquotedblleft Idea Heusler\textquotedblright, ERC
Advanced Grant No. 742068 \textquotedblleft TOPMAT\textquotedblright, EU FET Open RIA Grant No. 766566 \textquotedblleft ASPIN\textquotedblright, DFG through SFB 1143 (project-id 247310070) and the W{\"u}rzburg-Dresden Cluster of Excellence on Complexity and Topology in Quantum Matter – ct.qmat (EXC 2147, project-id 39085490).

\bigskip  
%%%%%%%%%%%%%%%%%%%%%%%%%%%%%%%%%%%%%%%%%%%%%%%%%%%%%%%%%%%%%%%%%%%%%%
%

\end{document}